\documentclass[prl,american,aps,twocolumn,showpacs,unsortedaddress]{revtex4}
\usepackage{graphics,bm}
\usepackage{amssymb}
\usepackage{epsfig}
\usepackage{epsfig}
\usepackage{epsf}
\usepackage{graphicx}
\usepackage{amsmath}
\usepackage{mathrsfs}
\usepackage{amsfonts}

\usepackage{xcolor}

\makeatletter

\newcommand{\beqa}{\begin{eqnarray}}
\newcommand{\eeqa}{\end{eqnarray}}
\newcommand{\beq}{\begin{equation}}
\newcommand{\eeq}{\end{equation}}

\begin{document}

\title{\bf{Quantum fluctuations   stabilize Skyrmion textures }}

\author{A. Rold\'an-Molina}
\affiliation{Instituto de F\'isica, Pontificia Universidad Cat\'olica de Valpara\'iso, Avenida Universidad 330, Curauma, Valpara\'iso, Chile}
\author{M. J.  Santander}
\author{\'A.  S. N\'u\~{n}ez} 
\affiliation{Departamento de F\'isica, Facultad de Ciencias F\'isicas y Matem\'aticas, Universidad de Chile, Casilla 487-3, Santiago, Chile}
\author{J.  Fern\'andez-Rossier\footnote{Permanent Address: Departamento de F\'isica Aplicada, Universidad de Alicante}} 
\affiliation{International Iberian Nanotechnology Laboratory, Av. Mestre Jose Veiga, 4715-310 Braga, Portugal}

\begin{abstract}
 Here we show that the zero point energy associated to the quantum spin fluctuations of a non-collinear spin texture produce Casimir-like magnetic fields. 
We study the effect of  these Casimir fields on the    topologically protected  non-collinear spin textures known as skyrmions. 
We calculate  the zero point energy, to the lowest order in the spin wave expansion,  in a  Heisenberg model with 
Dzyalonshinkii-Moriya interactions chosen so   that  the classical ground state displays skyrmion textures, that disappear upon application of a strong enough magnetic field. 
Our calculations show that the Casimir magnetic field contributes a 10 percent of the total Zeeman energy necessary to  delete the skyrmion texture with an applied field, and is thereby an observable effect.

\end{abstract}


\maketitle


The properties of many magnetic materials with broken spin symmetries are  correctly described in terms of magnetic moments that have a well defined orientation and are, in this sense, classical.  Yet quantum fluctuations, that ultimately arise from the fact that the different  projections of the spin operator do not commute with each other, are invariably present.  The standard\cite{Yosida,Auerbach}
description of spin excitations of these broken symmetry magnetic materials in terms of quantum spin waves implies that, even in the ground state, there is  a zero-point (ZP) energy ultimately associated to fluctuations of the magnetization field\cite{Yosida,Auerbach}.   This ZP energy vanishes in the case of collinear ferromagnets, but is non-zero in general.

In the case of quantum electrodynamics,  the vacuum energy, $\mathcal{E}_{ZP}$, associated to the photon zero point motion  stored between two parallel plates depends on the distance between $L$ between them, resulting on  an effective Casimir force\cite{Casimir} $F=-\frac{\delta \mathcal{E}_{ZP}}{\delta L}$. 
In this letter we show that, analogously,  
  the vacuum energy associated to the quantum spin fluctuations of a given spin texture depends on an ensemble of classical variables encoded in the local spin orientation $\vec{\Omega}_i$, resulting in an effective Casimir field  $\vec{\tau}_i=-\frac{\delta  \mathcal{E}_{ZP}}{\delta \vec{\Omega}_i}$.
Here we show that proper account of those fluctuations leads to a reduction in the energy of skyrmion textures rendering them stable in regions where a purely classical analysis predicts them to be unstable. 
  
Magnetic Skyrmions are topologically protected spin structures\cite{Skyrme}. Observed recently, both in chiral magnets\cite{Muhlbauer,Jonietz,Neubauer,Munzer,PMilde,Nagao,Seki} and in engineered surfaces\cite{Heinze,Romming2013}, they have received attention for potential applications in spintronics because it is possible to control their position with very low current densities\cite{Fert}. In addition skyrmions display several features that are convenient from the 
viewpoint of potential applications, such as dimensions within the nanometric scale, 
topological protection, and high mobility.  They have been observed in  bulk magnets MnSi\cite{Muhlbauer,Jonietz,Neubauer}, Fe$_{1-x}$Co$_x$Si\cite{Munzer,PMilde,Pfleiderer,Yu2}, Mn$_{1-x}$Fe$_x$Ge\cite{Shibata} and FeGe\cite{Yu1} using neutron scattering and Lorentz transmission electron microscopy. As it was reported, in these systems an external magnetic field can induce skyrmions with diameters of about a few tens of nanometers. 
The inclusion of spin transfer torques, as it is shown by numerical simulations, can be used to nucleate and manipulate isolated skyrmions\cite{Iwasaki,Iwasaki2}. In addition,  two dimensional atomic-scale magnetic nanoskyrmion lattices have also been observed by means of spin polarized scanning tunneling microscopy  in a monoatomic layer of Fe atoms on top of Ir(111) surface\cite{Heinze}. In this case, skyrmions arrays with a length scale of only one  nanometer are stabilized by the interplay between Dzyaloshinskii-Moriya 
interaction and the breaking of inversion symmetry at surfaces and interfaces. 

Previous work \cite{Batista} has addressed the study of skyrmions spin excitations in a semiclassical approximation, for rather large systems with up to $10^5$ localized moments. An effective theory of magnon-skyrmion scattering in a continuum approach has been recently carried out by Sch\"utte \textit{et al.} \cite{Garst}. Here we treat smaller skyrmions, using a complementary approach that permits to study quantum fluctuations associated to the excitations relevant for atomic scale skyrmions \cite{Heinze} for which  the effects of quantum fluctuations of the magnetization cannot be neglected at the outset.


%
%
%
The non-collinear  spin alignment between  first neighbors is promoted by the competition between the standard
Heisenberg interaction ($-J\vec{S}_{i}\cdot \vec{S}_{j}$) and the antisymmetric Dzyaloshinskii-Moriya interaction (DMI) term $\vec{D}_{i,j}\cdot\vec{S}_{i}\times \vec{S}_{j}$, which is only present in 
 non-centrosymmetric systems. In particular, this competition leads to helical spin textures with fixed chirality.
Application of a magnetic field $\mathbf{h}$ along the $\hat{z}$ direction (perpendicular to the system surface and then to $\vec{D}_{i,j}$ for any pair $i,j$)
breaks the up-down symmetry in the system resulting, for a range of fields, in an abrupt change in the ground state. In this manner, the external magnetic field and a uniaxial anisotropy can stabilize an isolated Skyrmion on a ferromagnetic system with DMI. A minimal model that captures the essence of the physics just described is given by the following Hamiltonian with four contributions, the Heisenberg and DM interactions, the Zeeman term and the uniaxial anisotropy:  
\begin{eqnarray}
H&=&-\sum_{<i,j>}J_{i,j}\vec{S}_{i}\cdot \vec{S}_{j}+\sum_{<i,j>}\vec{D}_{i,j}\cdot(\vec{S}_{i}\times \vec{S}_{j})\nonumber
\\
&-& \sum_{i}\vec{h}\cdot\vec{S}_{i}-K\sum_{i}(S_{i}^{z})^{2},
\label{Hamiltonian1}
\end{eqnarray}
where the interactions are taken  with nearest neighbors only.
 
Since exact diagonalization of this Hamiltonian is not possible except for very small systems,  we adopt
the standard spin wave approach, that is implemented it two clearly separated steps. First, we treat eq. (\ref{Hamiltonian1}) as a classical functional 
${\cal E}_{cl}\equiv H(\vec{\Omega}_i)$,
where the spin operators $\vec{S}$ are replaced by classical vectors $\vec{\Omega}_i$.  
The classical ground state, defined as the configuration $\vec{\Omega}_i$ that minimizes the functional ${\cal E}_{cl}$,  is determined by self-consistent iteration. 
The above Hamiltonian is studied for square systems with up to 45$\times$45 sites.  
We simulate hard boundary conditions at the edges by the inclusion of a large magnetic field in the border sites. 
Unless otherwise stated, we assume \cite{Batista,NagaosaDynamics} S = 1, $J$ = 3.0 meV, $D$ = 0.6 meV, $\vec{D}_{i,j}\parallel \vec{r}_{i,j}$ and $K = 0.5 (D^2/J)$, where $\vec{r}_{i,j}$ is the unit vector between the sites $i$ and $j$. 
 We find that, for this choice of parameters,   the   skyrmion configuration is more stable than the ferromagnetic state for fields up to $h_c\sim 0.8 (D^2/J)$  and skyrmion solutions can not be found at all for $h_{c'}\sim 1.67 (D^2/J)$ .  
Importantly,   the main results discussed below do not depend qualitatively on the choice of Hamiltonian parameters.
 

The second step permits to compute  quantum fluctuations of the classical solution. For that matter,
 we represent  of the spin operators
 in terms of Holstein Primakoff (HP) bosons\cite{HP,Auerbach}:
\begin{eqnarray}
\mathbf{S}_i\cdot\vec{\Omega}_i&=&S-n_i,\,\, \nonumber \\
  S_{i}^{+}&=&\sqrt{2S-n_{i}}\ a_{i}, 
 \,\,S_{i}^{-}=a^{\dagger}_{i}\sqrt{2S-n_{i}}\ ,
\end{eqnarray}
where $\vec{\Omega}_{i}$ is the spin direction of the classical ground state on the position $i$, $a_i^{\dagger}$ is a Bosonic creation operator and $n_i=a^{\dagger}_ia_i$ is the boson number operator.  The operator $n_i$   measures  the deviation of the system from the classical ground state.   The essence of the spin wave calculation is to 
represent  the spin operators in Eq. (\ref{Hamiltonian1}) by the HP bosonic representation and 
to  truncate the Hamiltonian up to quadratic order in the bosonic operators.  The linear terms in the bosonic operators vanish when the expansion is done around the correct classical ground state.   This approach has been widely used in the calculations of spin waves for ferromagnetic and antiferromagnetic ground states\cite{Yosida,Auerbach}. 
In the spin wave approximation we thus approximate $S_{i}^{-}=\sqrt{2S}\ a_{i}^{\dagger}$, so that the creation of a  HP  boson is equivalent the removal of one unit of spin angular momentum from the classical ground state.


 After some algebra\cite{ARM1} the quadratic Hamiltonian can be reduced to a symmetric form in creation an annihilation operators written as: 
\begin{eqnarray}
H=\mathcal{E}_{cl}+\mathcal{E}_0+\sum_{i,j} \left( a_{i}^{\dagger}, a_{j}\right) 
\left(
\begin{array}{cc}
t_{i,j} & \tau_{i,j}\\
\tau^*_{i,j} & t_{i,j}^*
\end{array}
\right)
\left(
\begin{array}{cc}
a_j \\ a^{\dagger}_j
\end{array}
\right)
\label{Hamiltonian2}
\end{eqnarray}     
where the constant $\mathcal{E}_0$  is a quantum contribution to the ground state energy that  arises naturally in this expression  as a consequence of applying the bosonic conmutation relations in order to bring the Hamiltonian to its symmetric form. The second and third term in eq. (\ref{Hamiltonian2}) correspond to the called spin waves Hamiltonian $H_{SW}$. 
Importantly,  $\mathcal{E}_{cl}$, $\mathcal{E}_0$, $t_{i,j}$ and $\tau_{i,j}$ are
a functional of the classical ground state configuration $\vec{\Omega}_i$ that satisfies $\frac{\delta\mathcal{E}_{cl}}{\delta \vec{\Omega}_j}=0$.

\begin{figure}
 \centering
\includegraphics[width=0.2\textwidth]{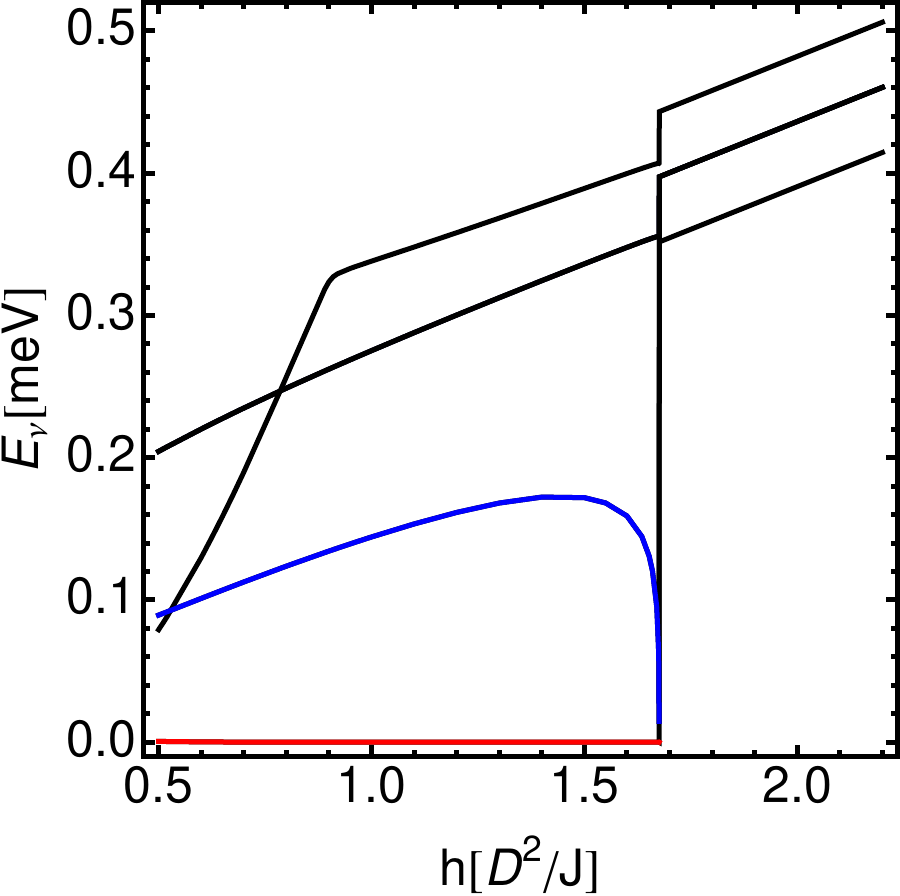}
  \includegraphics[width=0.215\textwidth]{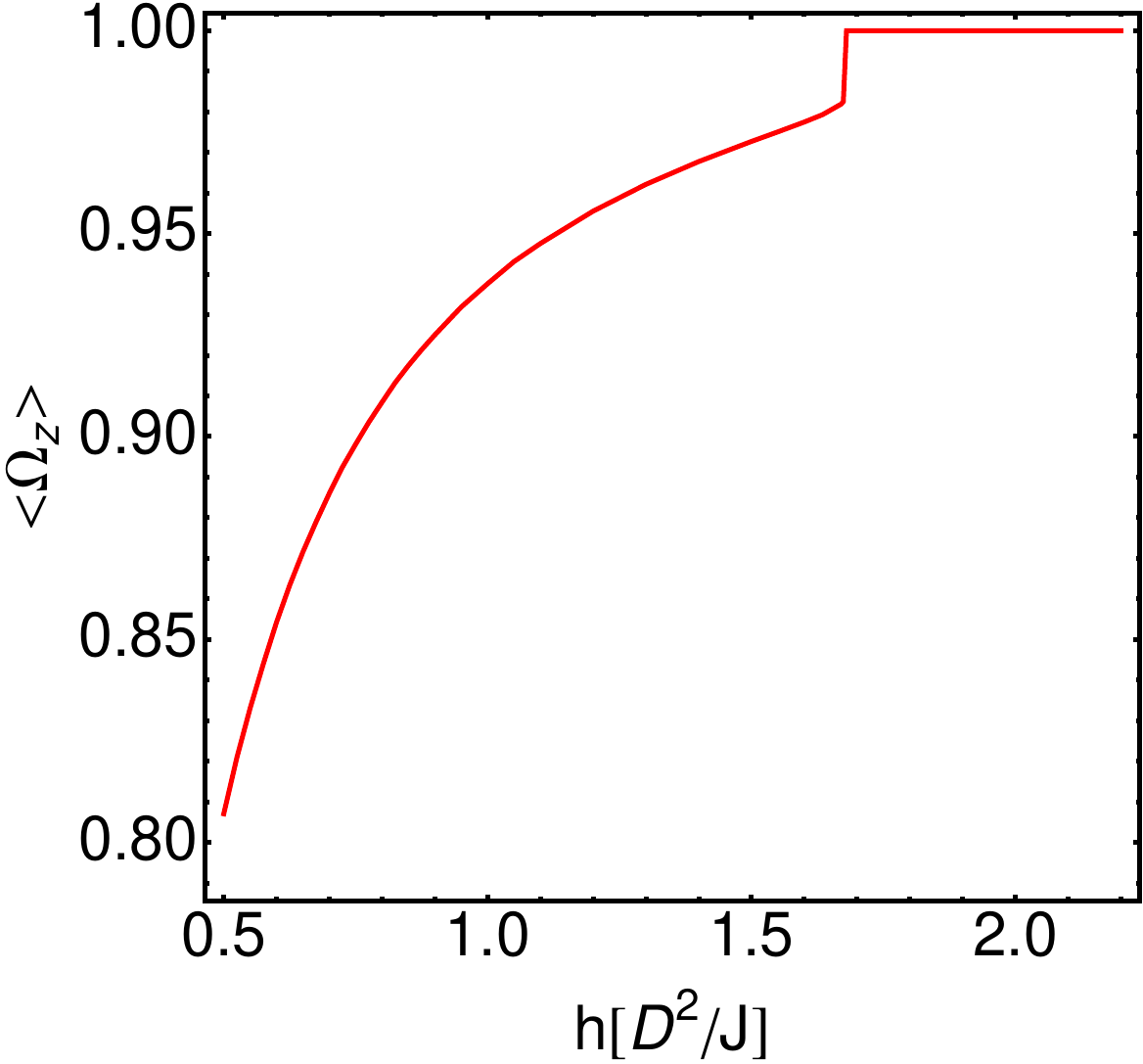}
\caption{  Left panel: Spin wave spectrum as a function of magnetic field. 
Right panel: Average spin projection along the $z$ axis as a function of the external magnetic field. As the field grows the size of the skyrmion is reduced. At a field of $h\sim 1.67 (D^2/J)$ the system can no longer sustain the skyrmion texture and a transition towards a ferromagnetic state is induced.  }
\label{fig: spectrumA05}
\end{figure}

\begin{figure}
 \centering
\includegraphics[width=0.55\textwidth]{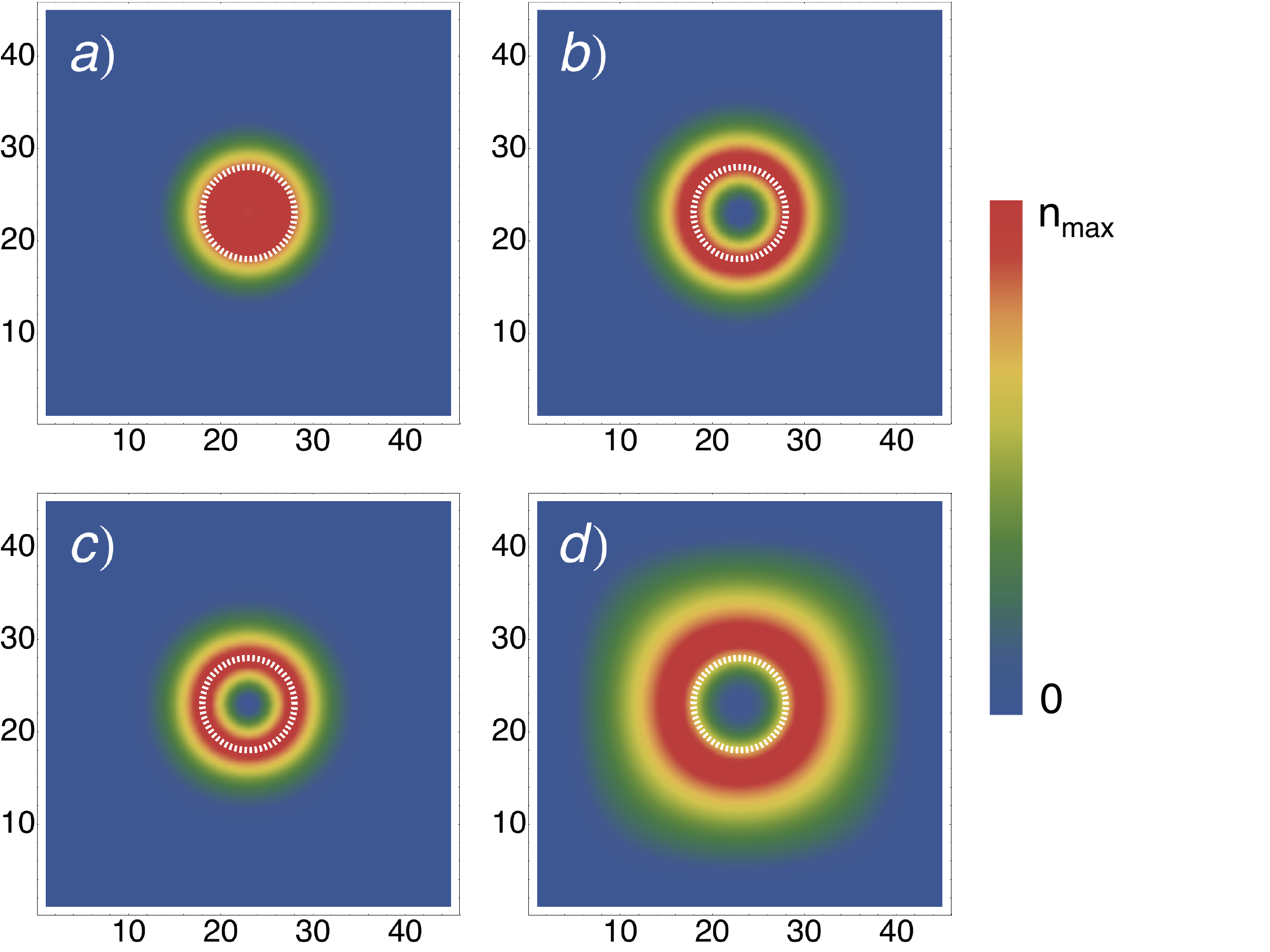}
\caption{Magnonic occupation of the four lowest energy excited states displayed in (a), (b), (c), and (d) respectively. The plots correspond to a field  $h\sim 0.75(D^2/J)$. On each figure the dashed white line correspond to the perimeter of the skyrmion. The maximum value of the occupation is $n_{max}=7.1\times 10^{-3}$ for the first excited state, $n_{max}=5.6\times 10^{-3}$ for the second excited state, $n_{max}=4.8\times 10^{-3}$ for the third excited state, and $n_{max}=2.1\times 10^{-3}$ for the fourth excited state.}
\label{fig: excited}
\end{figure}

The off-diagonal terms in eq. (\ref{Hamiltonian2}) do not conserve the number of HP bosons and  are in general present for non-collinear classical ground state  \cite{Yosida,Auerbach}. In consequence, the ground state of the Hamiltonian is not the vacuum of the $a$ bosons and, thereby, the ground state energy at the spin-wave approximation level is different from the classical ground state energy.   The Hamiltonian $H_{SW}$ is solved
 by means of a   paraunitary\cite{Colpa,ARM1} transformation, analogous to the usual Bogoliubov transformation for BCS Hamiltonians, that leads to:
 \begin{eqnarray}
H_{SW}=\mathcal{E}_0+\sum^N_{\nu=1}\hbar\omega_\nu (\alpha^\dagger_\nu	\alpha_\nu+\frac{1}{2})
\label{alphahamil}
\end{eqnarray} 
where the  $\alpha_{\nu}$ are bosonic  operators formed by linear combination of both $a$ and $a^{\dagger}$ that annihilate spin waves with energy  $\hbar \omega_\nu$.  In fig.\ref{fig: spectrumA05} we show the evolution of the four lowest energy modes as function of magnetic field , the abrupt change at 
$h_{c'}$ is a consequence of the fact that for larger fields, the  skyrmion like solution is no longer allowed and a transition to  ferromagnetic order takes place.  It is apparent that the skyrmion has in-gap spin wave excitations, in line with those obtained using the semiclassical approximation\cite{Batista}. The first excitation has a very low energy and can be readily associated with a gyrotropic traslation mode. In Fig. (\ref{fig: excited}) we show the magnonic occupation, defined as 
\begin{equation}
n^i_{\nu}=\langle \psi_{\nu}|a^{\dagger}_{i}a_{i}|\psi_{\nu}\rangle
\label{occupation}
\end{equation}
for the four lowest energy modes, where   $|\psi_{\nu}\rangle\equiv \alpha^{\dagger}_{\nu}|GS\rangle$. 
  
Inspection of the magnonic occupation in the four lowest energy excited states, shown in  Fig.\ref{fig: excited},
  suggest that the second and third correspond to a breathing modes.  The skyrmion-ferromagnet transition is driven by a reduction of the energy of the lowest energy breathing mode. The abrupt reduction of the second excitation energy near the critical field signals the instability of the Skyrmion texture. Below the critical field, both internal and extended spin wave states are present in the system. For fields larger than the critical field the spin wave spectrum correspond to the usual ferromagnetic spin waves.

We now  discuss  the main result of the manuscript.  In contrast with the semiclassical theory, the quantum theory of spin waves also yields a renormalization of the energy of  the ground state whose wave function is, by definition, 
the {\em vacuum} of the $\alpha$ operators,  $\alpha_{\nu}|GS\rangle=0$.  In the case of  non collinear classical order, the ground state is {\em  different} from the vacuum of HP bosons. The ground state  energy reads:
\begin{equation}
\mathcal{E}_{GS}=\mathcal{E}_{cl} +\mathcal{E}_0+\frac{1}{2}\sum^N_{\nu=1} \hbar\omega_\nu\equiv 
\mathcal{E}_{cl} +\mathcal{E}_{ZP}
\end{equation}
which is different from the classical energy due to the contribution coming from quantum fluctuations, or zero point energy, 
$\mathcal{E}_{ZP}\equiv \mathcal{E}_{GS}-\mathcal{E}_{cl} $.
In the case of ferromagnetic solutions it turns out that  $\mathcal{E}_{ZP}$ vanishes  \cite{Yosida,Auerbach}
because  the collinear ferromagnetic ground state is also an eigenstate of the  exact Hamiltonian. 

\begin{figure}[hbt]
 \centering
\includegraphics[width=0.5\textwidth]{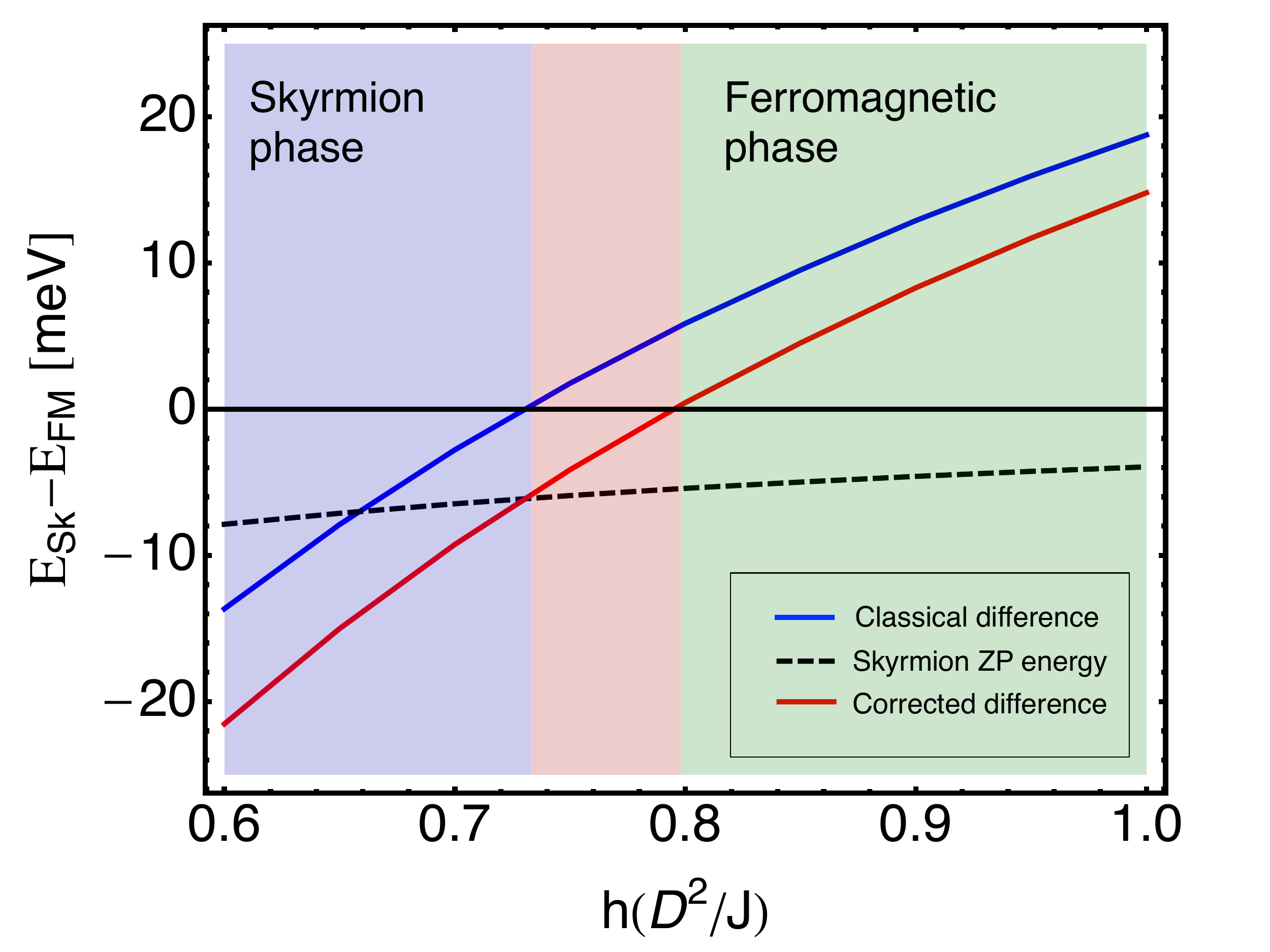}
\caption{Differences between skyrmion and ferromagnetic configuration energy. The crossing between the curves and the horizontal axis sets the boundary for the region of stability for the skyrmion in classical and corrected approaches (blue and red lines respectively). 
 Corrected energy difference between the skyrmion and ferromagnetic textures including the zero point fluctuations,
 enlarging the region of stability for   skyrmions.}
\label{fig: energy}
\end{figure}

\begin{figure}[hbt]
 \centering
\includegraphics[width=0.47\textwidth]{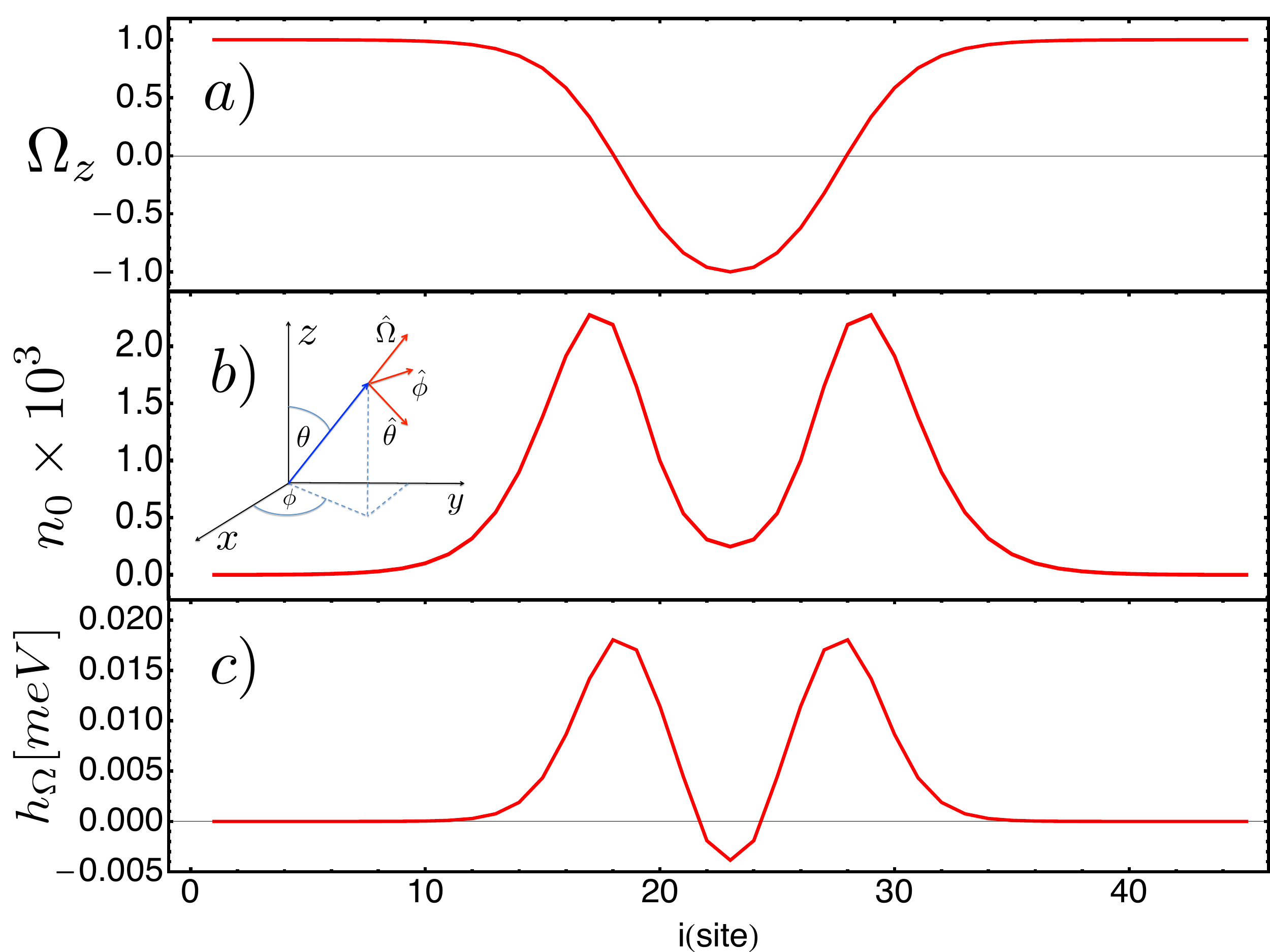}

\caption{Magnetization profile on $z$ direction (a), magnonic occupation (b) and Casimir field ($\Omega$ direction) (c) on a skyrmion ground state for h=0.75$(D^2/J)$. The local coordinate system for spins and the corresponding Casimir fields is shown in (b) inset. As consequence of the Casimir field on direction $\Omega$, the skyrmion zero point energy is reduced thereby it increase the texture stability.}
\label{fig: n0A05}
\end{figure}

 In general the vacuum energy $\mathcal{E}_{ZP}$  is not zero and, as it turns out,  it contributes significantly to the relative stability of different spin configurations.  We show this for the case of a skyrmion. In   Fig. \ref{fig: energy}  we compare the energy difference between the skyrmion and the collinear solution, as a function of the applied field $h$,  calculated at the classical level (blue line) and including $\mathcal{E}_{ZP}$ (red line). It is apparent that the critical field $h_c$ above which the collinear solution is more stable increased 10\%  when the $\mathcal{E}_{ZP}$ is included: the skyrmion is more stable due to quantum fluctuations. 
 By analogy with the Casimir effect, variations of the vacuum energy with the local moment orientation can be associated with an effective Casimir magnetic field.  This Casimir field  is the magnetic analogue of the Casimir force.
In this manner, the effective field on a given spin can be defined as: 
\begin{equation}
\vec{h}_{i}= -\frac{\delta \mathcal{E}_{GS}}{\delta \vec{\Omega}_i}=-\frac{\delta}{\delta \vec{\Omega}_{i}}\left(\mathcal{E}_{cl}+\langle GS\left| \mathcal{H}_{SW}\right|GS\rangle \right)
\end{equation}
The first contribution is the classical field, which vanishes identically by construction.  Using this and the Hellman-Feynman theorem\cite{Feynman},  the Casimir field $\vec{h}_{ZP}$ read: 
\begin{equation}
\vec{h}_{ZP}^{i}=-
\langle GS\left| \frac{\delta\mathcal{H}_{SW}}{\delta \vec{\Omega}_i}\right|GS\rangle
\end{equation}
and they can be readily calculated.  Our results for the Casimir field  are displayed in Fig. (\ref{fig: n0A05}), together with the projection of the classical magnetization over the $z$ axis (top panel), and the expectation value, computed with the  ground state wave function,  of the magnon occupation.  
For reference, we sketch the spherical coordinate system as shown in the inset of Fig. (\ref{fig: n0A05})(b). 
At the center of the skymion the magnetization points along $-\hat{z}$, opposite from the magnetization of the surroundings, as can be seen in Fig. (\ref{fig: n0A05})(a).

The stabilization of the skyrmion configuration with respect to the applied field can be associated to the Casimir field using the following argument. The Zeeman contribution enters the energy of the system as $-\Omega_z h$, in the same manner, the contribution of the Casimir field in classical direction $\vec{\Omega}$ correspond to $-S\ (\vec{h}_{ZP}\cdot \vec{\Omega})=-S\ h_{\Omega}$.
In Fig. (\ref{fig: n0A05})(c) we plot $h_{\Omega}$, it is clear from here that the contribution of the Casimir field on the classical direction is to lower the energy of the skyrmion state.




In addition to the stabilization of the skyrmion solution, quantum fluctuations can have other observable consequences that we discuss now. 
 In order to characterize the density of spin fluctuations of a given eigenstate of Hamiltonian (\ref{alphahamil}),
   $|\psi_{\nu}\rangle\equiv \alpha^{\dagger}_{\nu}|GS\rangle$,  
  we have defined the magnonic occupation in Ec.(\ref{occupation}).
Importantly, this quantity is non zero even in the ground state, $n^{i}_{0}=\langle GS|a^{\dagger}_{i}a_{i}|GS\rangle$, reflecting the zero point quantum fluctuations that are a consequence of a non collinear classical state.  
The zero-point magnonic occupation of the skyrmion ground state is shown in figure (\ref{fig: n0A05})(d).  Calculations using different values for $D,J,K$ show that  the magnitude of the ZP fluctuations increases as the size of the skyrmion is reduced.
The ZP fluctuations have observable consequences.   Since $\langle\vec{S}\cdot\Omega_i\rangle=S- \langle a^{\dagger}_{i}a_{i}\rangle$, both the zero point motion and the thermally excited spin waves will reduce the effective atomic magnetic moments of skyrmion solutions, an effect that should be observable with local probes, such as STM.   

%

%
%
%

Importantly, the Casimir magnetic field exert torque on the magnetic moments, the Casimir spin torque, defined as $\vec{\tau}_{i}=\vec{\Omega}_{i}\times\vec{h}_{ZP}^{i}$. The ZP energy, and thereby the ZP Casimir spin torques, depend on the energy of the spin wave excitations, which are interesting on their own right and could be observed by means of inelastic electron tunneling spectroscopy, as recently demonstrated for spin waves in short ferromagnetic chains\cite{Spinelli2014}.


%

%
%
%
%
%
%

 In summary, we have studied the observable consequences of quantum spin fluctuations in skyrmions. We have found that the zero-point energy stabilizes  skyrmions  with respect to the application of  a magnetic field, increasing
 significantly the critical field necessary to ears  the skyrmion.     We show that this stabilization can be associated
 to a Casimir magnetic field driven by zero point fluctuations analogous to the Casimir effect.
 We have also found that zero point fluctuations renormalize the shape of the skyrmion and shrink the atomic magnetization, an effect that becomes significant for small skyrmions and could be observerved by means of spin-polarized scanning tunneling microscopy. 

The authors would like to thank funding from grants Fondecyt 1150072, ICM P10-061-F by Fondo de Innovaci\'on para la Competitividad-MINECON and Anillo ACT 1117. ASN also acknowledges support from Financiamiento Basal para Centros Cient\'ificos y Tecnol\'ogicos de Excelencia, under Project No. FB 0807(Chile).  ARM,  MJS and ASN  acknowledge hospitality of INL. 

\end{document}